\documentstyle[12pt,aaspp4]{article}
\begin{document}

\title{Monitoring All Sky for Variability}

\author{Bohdan Paczy\'nski}
\affil{Princeton University Observatory, Princeton, NJ 08544--1001, USA}
\affil{E-mail: bp@astro.princeton.edu}

\begin{abstract}
A few percent of all stars are variable, yet over 90\% of variables brighter
than 12 magnitude have not been discovered yet.  There is a need for an all
sky search and for the early detection of any unexpected events: optical 
flashes from gamma-ray bursts, novae, dwarf novae, supernovae, `killer 
asteroids'.  The ongoing projects like ROTSE, ASAS, TASS, and others, using
instruments with just 4 inch aperture, have already discovered thousands of 
new variable stars, a flash from an explosion at a cosmological distance, and
the first partial eclipse of a nearby star by its Jupiter like planet.  About
one million variables may be discovered with such small instruments, and many
more with larger telescopes.  The critical elements are software and full 
automation of the hardware.  A complete census of the brightest eclipsing 
binaries is needed to select objects for a robust empirical calibration of the
the accurate distance determination to the Magellanic Clouds, the first step
towards the Hubble constant.  An archive to be generated by a large number of
small instruments will be very valuable for data mining projects.  The real
time alerts will provide great targets of opportunity for the follow-up
observations with the largest telescopes.
\end{abstract}

\keywords{galaxy: structure -- stars: binaries: eclipsing -- stars: cepheids 
-- stars: evolution -- stars: variables -- techniques: photometric}

\section{Introduction}

The aim of this paper is to present an outline of the idea which, I expect,
will lead to the development of a series of long term observing projects,
with the general goal to monitor all sky for variability.  This concept is
not new, and I am not in a position to present a full historical background.
However, it is well known that the projects of this kind were carried out
in the past and resulted in a magnificent archive of several hundred thousand
photographic plates at Harvard and elsewhere.  There are several on going
projects which image all sky visible from a particular site every clear night,
and archive the data, for example ROTSE and LOTIS.  There are several other
projects under development, and in various stages of implementation, like
ASAS, STARE and TAROT.  A lot of information about these and many other 
projects is available on the World Wide Web.  The links can be found on my 
home page\footnote{http://www.astro.princeton.edu/faculty/bp.html}.  This paper
has been influenced by my experience with the OGLE (Udalski et al. 1997
\footnote{http://www.astro.princeton.edu/\~~ogle/}) and ASAS (Pojma\'nski 2000
\footnote{http://archive.princeton.edu/\~~asas/}), and to a large
extent it presents my plans for the future.

The existing catalogs of variable stars are very incomplete.  Pojma\'nski (2000)
discovered almost 4,000 variable stars brighter than $ I = 13 $ mag in a small
area of the sky, just 300 square degrees.  Only 4\% of these are listed in any
catalog of variable stars, while 96\% are new discoveries.  Using archive
ROTSE data Akerlof et al. (2000) found 1781 periodic variable stars brighter
than 14 mag in 2,000 square degrees, 90\% of these were new discoveries.  All
these observations were made with very small, automated instruments, with an
aperture of only 4''.  There may be about one million variable stars in the
whole sky accessible to ROTSE and ASAS, waiting to be discovered.

Why should we bother?  The simplest answer is: because those variables are
out there.  For those who would like to know more specific reasons I shall
list some in the next section, also see Paczy\'nski (1997).

How should the searches be done?  There are many approaches.  Some are very
focused, like ROTSE and LOTIS, concentrating on a single goal: to detect
optical flashes associated with gamma-ray bursts.  In addition to a large
number of upper limits there was one spectacular discovery of an optical
flash associated with the GRB 990123, which peaked at 9 mag (Akerlof et al.
1999), with the source being at a large redshift, $ z = 1.6 $.  By design
ROTSE archives the data, but a broad search of variability is not high 
on the priority list, though it has generated an interesting paper (Akerlof 
et al. 2000).  Other projects, like VSNET, monitor variability of a large 
number of preselected stars.  Still other projects, like ASAS, are not focused
on any particular type of variability or any specific objects, 
with a broad goal to monitor everything that varies in brightness
(stars, AGNs, etc.) or position
(asteroids).  Preliminary results were presented by Pojma\'nski (1997, 1998,
2000).  In this paper I am making a case for such broad searches.

How should the projects evolve?  I think it makes sense to start small and to
expand gradually.  There are several reasons.  A small projects costs less
than a big one, it is easier to develop, and a relatively small data
rate is easier to handle.  Software, which is the most difficult and
expensive part of any survey, can be developed gradually.  Once a small
system works, all the way from photons entering the aperture to the first
paper(s) with the early results, it is relatively easy to expand, having learnt
what are the difficult elements.  This is the pattern followed by the OGLE
and ASAS. 

There are several important aspects of data handling: real time data
processing, calibration, developing 
public domain archives and alerts.  Problems in these areas are as much
technological as sociological or psychological.  To take full advantage
of the all sky monitoring an easy public access to the data is very important.

\section{The Goals}

The many goals of all sky variability searches were listed and discussed
in the past (Paczy\'nski 1997).  The following is a brief summary.

A well calibrated survey with well described search method will provide
large complete samples of eclipsing binaries,
pulsating stars, exploding stars, stars with large proper
motions, quasars, asteroids, comets, and other types of objects.
Complete samples are important for statistical studies of the galactic
structure, the stellar evolution, the history of our planetary system, etc.  

A large number of variables of any specific type will allow future selection
of the `cleanest', simplest objects.  A physicist in his or her laboratory 
may create conditions which allow a study of a specific phenomenon with
a minimum of disturbances.  An astronomer cannot influence the universe,
various objects can be only observed.  However, given a large
number of objects an astronomer may select one (or several) which are the
simplest, the `cleanest' from some particular point of view, and analyze them
in great details with follow-up observations, which will provide a better 
understanding and/or a better calibration of the particular type of objects 
and/or processes.

In a large sample some very rare objects
or events will be detected.  A spectacular
example from the past is FG Sagittae, a nucleus of a planetary
nebula undergoing a helium shell flash in front of our telescopes
(Woodward et al. 1993, and references therein).
Some may provide spectacles which bring astronomy to the general public.
Some recent examples are the supernova 1987A in the Large
Magellanic Cloud, and a collision of the comet Shoemaker-Levy
with Jupiter in the summer of 1994.

Fully automated real time data processing will provide
instant alerts for a variety of unique targets of opportunity: optical
flashes from gamma-ray bursts, novae, dwarf novae,
supernovae, gravitational microlensing events, small asteroids
that collide with earth every year, etc.  Such alerts will provide
indispensable information for the largest and most expensive space
and ground based telescopes, which have tremendous light collecting 
power and/or resolution but have small fields of view.

The archive of photometric measurements will provide
a documentation of the history for millions of objects, some of which
may turn out to be very interesting in the future.  The
Harvard patrol plates and the Palomar Sky Survey atlas provide an
excellent example of how valuable an astronomical archive can be.

New objects and phenomena will be discovered; this is virtually guaranteed
whenever the volume of data increases by several orders of magnitude.

A specific example of targets for calibration to be provided by the
all sky searches are the detached eclipsing binaries.  The distance to
the Magellanic Clouds is the important step towards the determination of
the Hubble constant, and its uncertainty is the largest contributor to
the error in the $ H_0 $ obtained by the HST Key Project (see Mould et
al. 2000, and references therein).  The most direct and accurate distance
to the LMC and SMC obtainable with current technology is a century old
method based on detached eclipsing binaries.  An excellent outline of the
method with a complete list of historical references and the up-to-date
status is provided by Kruszewski \& Semeniuk (1999).  For the method to
be fully trusted it has to be calibrated with relatively nearby, and hence
bright eclipsing binaries, towards which purely geometric distances can be
measured by means of parallaxes or a combination of astrometry and radial
velocity amplitudes.  To select the best object for the task a complete
list of such systems is needed, yet the existing catalogs are very incomplete
(see Paczy\'nski 1997, Pojma\'nski 2000).

\section{Alerts and Archives}

It is somewhat surprising that while there are several examples of various 
alert systems that work well (e.g. microlensing alerts by MACHO in the past,
OGLE, and to some extent EROS in the past as well as now), large
archives of variability are not so easy to find.  There are technical as
well as sociological or psychological problems.  While my experience is
mostly based on microlensing searches, similar patterns can be found 
elsewhere.  

Some projects, like OGLE, try to release as much data as possible.
The reason: there is far too much data for a small team 
to fully analyze, and the more citations there are to the OGLE, the 
better it is for the project.  Still, only a small fraction of the total
has been released as quality control, calibration etc. are very labor
intensive and time consuming.  Microlensing photometry is placed on the Web
in near real time, but catalogs of pulsating and eclipsing stars take at least
a year to prepare, while there has been no general release of $ \sim 10^7 $
photometric measurements of $ \sim 10^5 $ variables of many kinds.
VSNET provides data for many interesting variable stars at their Web site
\footnote{http://www.kusastro.kyoto-u.ac.jp/vsnet/}.
Another example of almost immediate release of data is provided by the All Sky
Monitor team on the Rossi X-ray Timing Explorer
though the volume is relatively modest and hence easy to handle.

I expect that before too long robust enough software will be developed
to allow virtually all data from projects like OGLE and ASAS to be put
on the Web in near real time.  The question remains: how many teams will
be willing to follow this policy?  I think it depends on the way the
community evaluates the performance of a project, and on the policy according
to which tenure positions are filled.  It will be difficult to persuade
many to make their data instantly available to the public if the intellectual
effort and ingenuity needed for developing a fully functional system (hardware
and software) will be considered to be less valuable than the `science' of
plotting quantity `Y' versus quantity `X' and discovering a correlation
in somebody's well calibrated data.

A related issue is: should a survey project be broken down into the well
defined elements, each done by a single person or a small group? Or should
the whole effort be combined in a very large team, with all papers having
several dozen authors listed alphabetically, and no way to find out
whom to credit and whom to blame for different parts of the project?  It
is rare that more than a few people do the real work on which a particular
paper is based.  Is it more sensible to put all the names together, or
to cite as separate papers the well defined parts of the whole project?  
Note that the OGLE has only 6 or 8 members on its team, while ASAS is a
single person operation.  A very successful DUO project (Alard 1996a,1996b)
was mostly a work of a single graduate student.  
Variablity monitoring can also be done by amateur astronomers,
as demonstrated by AAVSO, and recentky by TASS (Richmond et al. 2000).
There is nothing 
intrinsic to all sky variability searches that requires huge teams
and a near anonymity of the real doers.   
Can the division of labor, with a proper recognition of
the diverse contributions, be implemented in astrophysics?
It works in economy: 
`The greatest improvement in the productive powers of labour,
and the greater part of the skill, dexterity, and judgment with
which it is anywhere directed, or applied, seem to have been the
effect of the division of labour' (Smith 1776).

I think that in a matter of just a few years it will be possible to have an
unrestricted access over the Internet to the up-to-date information about
the status of any bright variable star, as well as its past variability, with
the due credit being automatically given to those who will have provided the
data.  It is likely that only some people and some teams will follow this open
policy, but this may be sufficient to make it practical and to see what
impact will it have on astronomy in general.  I expect, or at least I hope,
that major contributors to the data archives and the alerts will be respected
enough to be offered tenure jobs at the major universities.

\acknowledgments{
This work was supported by the NSF grants AST-9819787 and AST-9820314.}



\begin{references}

\reference{} Akerlof, C., Balsano, R., Barthelmy, S., Bloch, J., Butterworth, 
P. et al.  1999, Nature, 398, 400

\reference{} Akerlof, C., Amrose, S., Balsano, R., Bloch, J., Casperson, D.
et al.  2000, AJ, 119, 1901

\reference{} Alard, C. 1996a, ApJ, 458, L17

\reference{} Alard, C. 1996b, IAU Symp. 173, p. 215

\reference{} Kruszewski, A. \& Semeniuk, I. 1999, Acta Astronomica, 49, 561

\reference{} Mould, J. R., Huchra, J. P., Freedman, W. L., Kennicutt, R. C.,
Ferrarese, L. et al. 2000, ApJ, 529, 786

\reference{} Paczy\'nski, B. 1997, proceedings of 12th IAP Colloquium: 
`Variable Stars and the Astrophysical Returns of Microlensing Searches', 
Paris (Ed. R. Ferlet), p. 357

\reference{} Pojma\'nski, G. 1997, Acta Astronomica, 47, 467 = astro-ph/9712146

\reference{} Pojma\'nski, G.  1998, Acta Astronomica, 48, 35 = astro-ph/9802330 

\reference{} Pojma\'nski, G.  2000, Acta Astronomica, 50, 177 = astro-ph/0005236

\reference{} Richmond, M. W., Droege, T. F., Gombert, G., Gutzwiller, M., 
Henden, A. A.  et al. 2000, PASP, 112, 397

\reference{} Smith, A. 1776, `The Wealth of Nations', Book I, Chapter I, the
  first sentence

\reference{} Udalski, A., Kubiak, M., \& Szyma\'nski, M. 1997, AcA, 47, 319

\reference{} Woodward, Ch. E. et al. 1993, ApJ, 408, L37

\end{references}
\end{document}